\documentclass[aps,prd,twocolumn,showpacs,eqsecnum,nofootinbib,floatfix]{revtex4-1}

\pdfoutput=1
\usepackage{amssymb}
\usepackage{gensymb}
\usepackage{graphicx}
\usepackage{amsmath}
\usepackage{wasysym}

\def\apjs{Astrophys.\ J.\ Supp.}
\def\apj{Astrophys.\ J.}

\def\apjl{Astrophys.\ J.\ Lett.}
\def\aap{Astron.\ \& Astrophys.}

\def\nat{Nature}
\def\prd{Phys.\ Rev.\ D}
\def\pasp{Pub.\ Astron.\ Soc.\ Pacific}

\begin{document}
\title{Morphological Tests of the Pulsar and Dark Matter Interpretations\\ of the WMAP Haze}
\author{J.\ Patrick Harding}
\email{hard0923@umd.edu}
\affiliation{Maryland Center for Fundamental Physics, Department of Physics, University of Maryland, College Park, MD 20742-4111}
\author{Kevork N.\ Abazajian}
\email{kev@umd.edu}
\affiliation{Maryland Center for Fundamental Physics, Department of Physics, University of Maryland, College Park, MD 20742-4111}
\date{\today}

\begin{abstract}
The WMAP haze is an excess in microwave emission coming from the center of the Milky Way galaxy. In the case of synchrotron emission models of the haze, we present tests for the source of radiating high-energy electrons/positrons. We explore several models in the case of a pulsar population or dark matter annihilation as the source. These morphological signatures of these models are small behind the WMAP Galactic mask, but are testable and constrain the source models. We show that detailed measurements of the morphology may distinguish between the pulsar and dark matter interpretations as well as differentiate among different pulsar models and dark matter profile models individually. Specifically, we find that a zero central density Galactic pulsar population model is in tension with the observed WMAP haze. The Planck Observatory's greater sensitivity and expected smaller Galactic mask should potentially provide a robust signature of the WMAP haze as either a pulsar population or the dark matter. 
\end{abstract}


\maketitle
\section{Introduction}
The Wilkinson Microwave Anisotropy Probe (WMAP) has provided a detailed map of the cosmic microwave background (CMB) as well as the foreground from our Galaxy~\cite{Hinshaw08,Gold08}. The foreground emission from the center of the Galaxy shows an excess of emission in the inner $5-20\degree$ around the Galactic center. This excess of microwave emission is known as the ``WMAP haze''~\cite{Finkbeiner04}. Due to the Galactic plane, the current data on the haze only goes down to about 6 degrees from the Galactic center. Two notable features of the haze are the approximate spherical symmetry and the strong angular dependence of the flux. In particular, the flux increases quickly toward the Galactic center.

The size and shape of the haze is currently not well constrained. Using a different foreground model may change the look of the haze slightly. For instance, Ref.~\cite{Bottino08} used a different synchrotron map to model the WMAP data and found that the haze in that case is smaller in spatial extent and slightly smaller in magnitude. A cleaner foreground subtraction is needed to further constrain the source of the haze. In this paper, we will use the measurement of the haze presented in Ref.~\cite{Hooper07}.

An explanation for the WMAP haze is a previously unknown source of microwave emission. The frequency dependence of the haze is quite hard, so a hard source like synchrotron radiation is a likely candidate~\cite{Finkbeiner07,Dobler07,Hooper07}. The magnetic field in the Galactic center tends to be tens of microgauss. Therefore, the synchrotron radiation from highly relativistic electrons and positrons near the Galactic center could be the source of the microwave signal~\cite{Ferriere09}. Another possibility for the source of the haze was thermal bremsstrahlung from ionized gas~\cite{Finkbeiner04}. However, the H$\alpha$ skymap shows no significant increase in the regions where the haze is strongest. At high density and high temperature ($\sim 10^{5}\rm K$) such a gas could still explain the haze, but the emission from such regions is constrained to be small, ruling it out as the source of the haze~\cite{Kurtz94}.

In the synchrotron emission interpretation, calculating the complete propagation of electrons in the interstellar medium requires the full diffusion-loss equation. This includes spatial diffusion of the $e^{+}e^{-}$, reacceleration of the particles due to momentum-space diffusion, energy loss due to several different mechanisms, convection of the particles in the Galaxy, and the particle source. To solve this complete equation, one may use the software GALPROP~\cite{Strong98,Strong00}. However, the major necessary physical features of the source and synchrotron emission can be sufficiently modeled using a Green's function solution to the diffusion-loss equation, which we employ~\cite{Baltz04,Baltz98}.

The source of high energy electrons and positrons required for synchrotron emission remains a mystery~\cite{Zhang08}. One intriguing explanation of the haze is from dark matter annihilations in the Galactic center~\cite{Finkbeiner07}. The $e^{+}e^{-}$ produced from by the annihilation products move through the Galactic magnetic field, creating the diffuse synchrotron emission~\cite{Hooper07}. This model matches the haze well, especially the strong angular dependence and approximate spherical symmetry.

It should be noted that Ref.~\cite{Cumberbatch09} claims that dark
matter annihilations cannot explain the haze because it would require
too large of a clumpiness boost factor. In this work, we employ
similar methods to those used in that paper and do not find that too
large of a boost factor is necessary. Ref.~\cite{Cumberbatch09}
averaged the dark matter density over the direction perpendicular to
the Galactic plane, which is inaccurate in light of the morphological
effects we present below.

Another source for the haze is $e^{+}e^{-}$ coming from the magnetosphere boundary of Galactic pulsars~\cite{Kaplinghat09}. This type of emission from nearby pulsars can be the source of positrons in the PAMELA results~\cite{Adriani08,Boulares89,Hooper08,Yuksel08,Profumo08} and also the source of features seen by ATIC~\cite{Chang08,Yuksel08,Profumo08} and the Fermi Gamma-Ray Space Telescope~\cite{Abdo09,Grasso09}.

In this paper, we explore models of the WMAP haze in the context of synchrotron radiation from the electron and positron production of Galactic pulsars and dark matter annihilations. We will show tests of this diffuse synchrotron emission which could distinguish between the exotic explanation of dark matter annihilations and the astrophysical pulsar sources. We also show how haze observations can test pulsar and dark matter models separately. These tests can be done with upcoming results from the Planck Observatory, which has been shown, e.g.\ in Leach, et al.~\cite{Leach08}, to be expected to have a much smaller Galactic mask than WMAP~\cite{Planck06,Stolyarov01}.

Recently, Kaplinghat et al.~\cite{Kaplinghat09} showed that the morphology of the haze is different depending on the source. This work pointed out that for a source with spherical symmetry, the haze should be slightly elliptically stretched along the Galactic plane while for a centrally-peaked pulsar source, the signal lies primarily along the Galactic plane and has less spherical symmetry. 

In this work, we examine the morphological structure of the WMAP haze in detail. First, we consider the signal due to pulsars, both from a Gaussian distribution and one which vanishes at the Galactic center. Second, we look at dark matter annihilations models' sensitivity to the dark matter density profile. Lastly, we compare the pulsar and dark matter scenarios to quantitatively distinguish them from one another. 
\section{Calculation of Synchrotron Flux}
To calculate the propagation of the electrons from their source, we must solve the diffusion-loss equation for $e^{+}e^{-}$:~\cite{Baltz04,Baltz98}
\begin{eqnarray} \label{diffloss}
\frac{\partial}{\partial t}\frac{dn}{d\epsilon}&=&\vec{\nabla}\cdot\left[K\left(\epsilon,\vec{x}\right)\vec{\nabla}\frac{dn}{d\epsilon}\right] \\ 
& &+\frac{\partial}{\partial\epsilon}\left[b\left(\epsilon,\vec{x}\right)\frac{dn}{d\epsilon}\right]+Q\left(\epsilon,\vec{x}\right)\ , \nonumber
\end{eqnarray}
with $K$ the diffusion constant, $b$ the energy loss rate, $Q$ the source term, and $\epsilon=E/\left(1 \rm\ GeV\right)$. This simplified diffusion-loss equation is derived from the full equation by assuming that reacceleration is small compared to spatial diffusion and that convection is small compared to the particle velocities. For highly relativistic $e^{+}e^{-}$ in the Galaxy, these are both reasonable assumptions.

For the simplified diffusion-loss equation, we shall use the parameterization of Ref.~\cite{Hooper07}. We assume a spatially-independent diffusion constant $K\left(\epsilon\right)=10^{28}\epsilon^{0.33}\rm\ cm^{2}s^{-1}$ inside a ``diffusion zone'' of half-thickness 3 kpc away from the Galactic plane and ``large'' in the radial direction of the Galactic plane. Outside this region, it is assumed that all particles are no longer confined to the Galactic magnetic field and therefore are free-streaming. Inside this region, the Galactic field is assumed to be a constant. This is a reasonable approximation because it is known that the Galactic magnetic field decreases quickly away from the Galactic center, but its exact profile is unknown.
The energy loss rate is also assumed to be spatially-independent and is parameterized by $b\left(\epsilon\right)=\epsilon^{2}/\tau_{E}$ with $\tau_{E}=2\times 10^{15}\rm\ s$. This characteristic energy-loss time $\tau_{E}$ accounts for losses due to inverse Compton scattering on the CMB and starlight, as well as synchrotron energy losses. The source term $Q\left(E,\vec{x}\right)$ is the rate of particles created per unit time per unit volume per unit energy. We will further assume that the diffusion-loss equation is in steady-state, so all terms are time-independent. The assumptions of spatial and temporal independence of the diffusion and loss parameters should be valid as long as we are concerned only with regions close to the Galactic center, so the average values of parameters can change only slightly in the time and distances involved. The assumption of time-independence in the source term assumes that the rate of pulsar formation and death is steady over the timescale of the electron propagation, during which only a few new pulsars should have formed. 

For our parameters, an analytic solution to the diffusion-loss equation can be found using Green's function techniques~\cite{Baltz04,Baltz98}. Writing the source term as $Q(E,\vec{x})=f(E)g(\vec{x})$, this analytic solution is given by:
\begin{equation} \label{dnde}
\frac{dn}{dE}=E^{-2}\int_{E}^{\infty}dE'f(E)\tau_{D}\left(E',E,\vec{x}\right)\ ,
\end{equation}
\begin{widetext}
\begin{eqnarray} \label{taud}
\tau_{D}(E', E, \vec{x})&=&\tau_{E}{\left(\pi D^{2}\right)^{3/2}}\sum_{n=-\infty}^{\infty}\left(-1\right)^{n}\int_{0}^{\infty}ds's'\int_{-L}^{L}dz'\int_{0}^{2\pi}d\theta'g(s',z') \\ 
&&\times\exp\left(-\frac{s^{2}+s'^{2}}{D^{2}}\right)\exp\left(\frac{2s s'\cos(\theta -\theta')}{D^{2}}\right)\exp\left(-\frac{\left(z-(-1)^{n}z'-2nL\right)^{2}}{D^{2}}\right)\ , \nonumber
\end{eqnarray}
\end{widetext}
where $\vec{x}=(s,\theta,z)$ are cylindrical coordinates from the Galactic center, $L=3\rm\ kpc$ is the end of the diffusion zone and $D^{2}=8.4\left(\epsilon^{-0.67}-\epsilon'^{-0.67}\right)\rm\ kpc^{2}$ is the integrated energy-dependent diffusion parameter.
To convert this $e^{+}e^{-}$ density to a synchrotron flux at earth, we must integrate over line-of-sight $\ell$ and then convolve it with the synchrotron spectrum for a single electron~\cite{Baltz04}:
\begin{equation} \label{flux}
\Phi\left(\alpha,\psi\right)=\frac{m_{e}^{2}c^{4}}{4\pi\tau_{\rm syn}\nu_{B}}\int d\epsilon \int d\ell\frac{dn}{dE}\left(\frac{1}{P}\frac{dP}{dy}\right)
\end{equation}
\begin{equation} \label{synchspec}
\frac{1}{P}\frac{dP}{dy}=\frac{27\sqrt{3}}{32\pi}y\int_{0}^{\pi}d\phi\sin(\phi)\int_{y/\sin(\phi)}^{\infty}K_{5/3}(x)dx \ .
\end{equation}
Here, $\tau_{\rm syn}=4\tau_{E}$ is the energy loss time due to synchrotron emission, $y=(\nu m_{e}^{2}c^{4})/(\nu_{B}E^{2})$, $\nu$ is frequency, and $\nu_{B}=(3eB)/(4\pi m_{e})$ is the characteristic synchrotron frequency. As in Ref.~\cite{Hooper07}, we choose a constant magnetic field of $10\ \mu G$ in the Galactic center and look primarily at the $\nu=22\rm\ GHz$ band of the haze. We will use a coordinate system where $\alpha$ is the angle out from the Galactic center and $\psi$ is the angle up from the Galactic plane.
\section{Electron Source Function}
\subsection{Source Function for Pulsars}
\begin{figure*}[t]
\begin{center}$
\begin{array}{cc}
\includegraphics[width=3.2in]{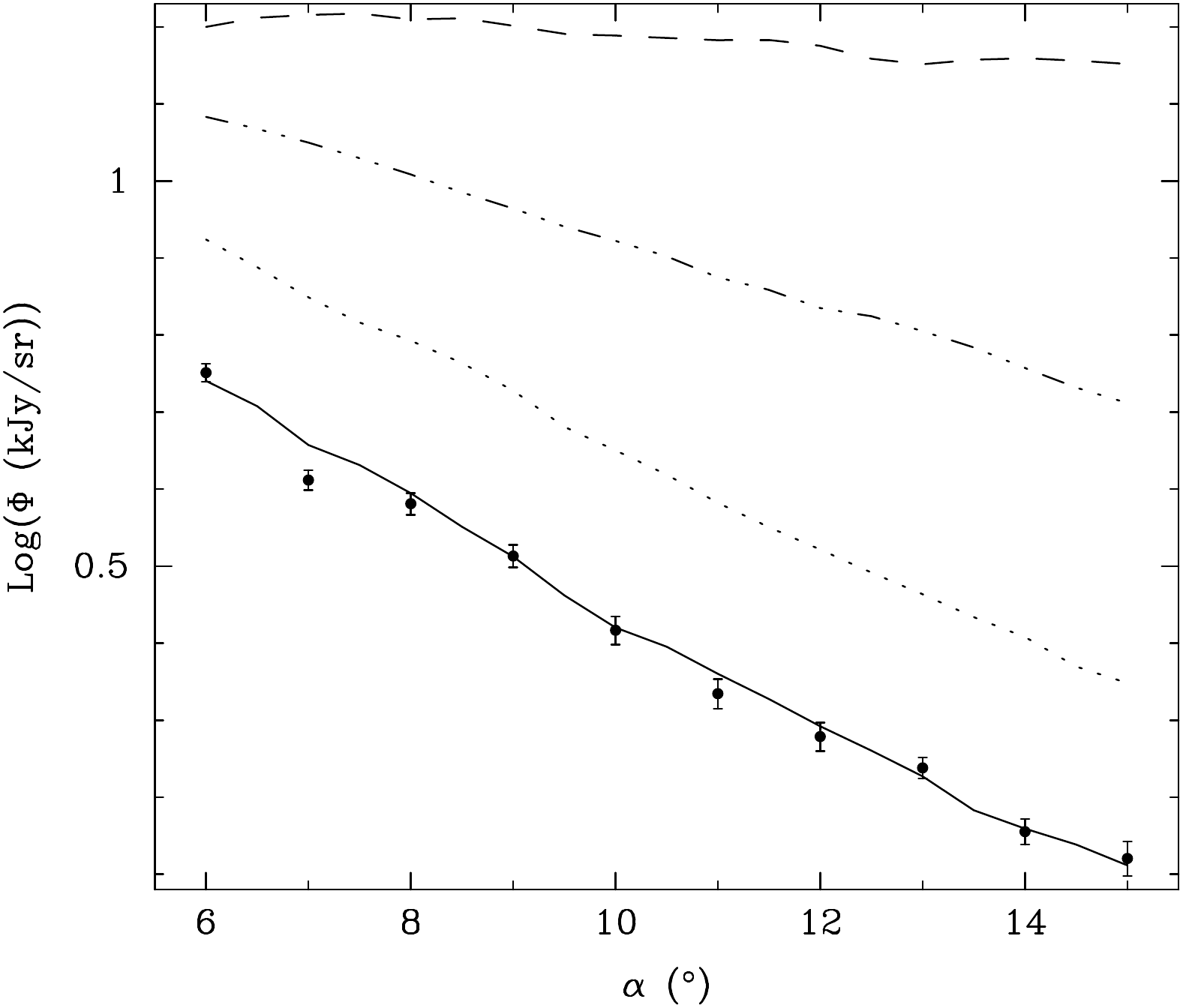} &
\hspace{1.0cm}
\includegraphics[width=3.2in]{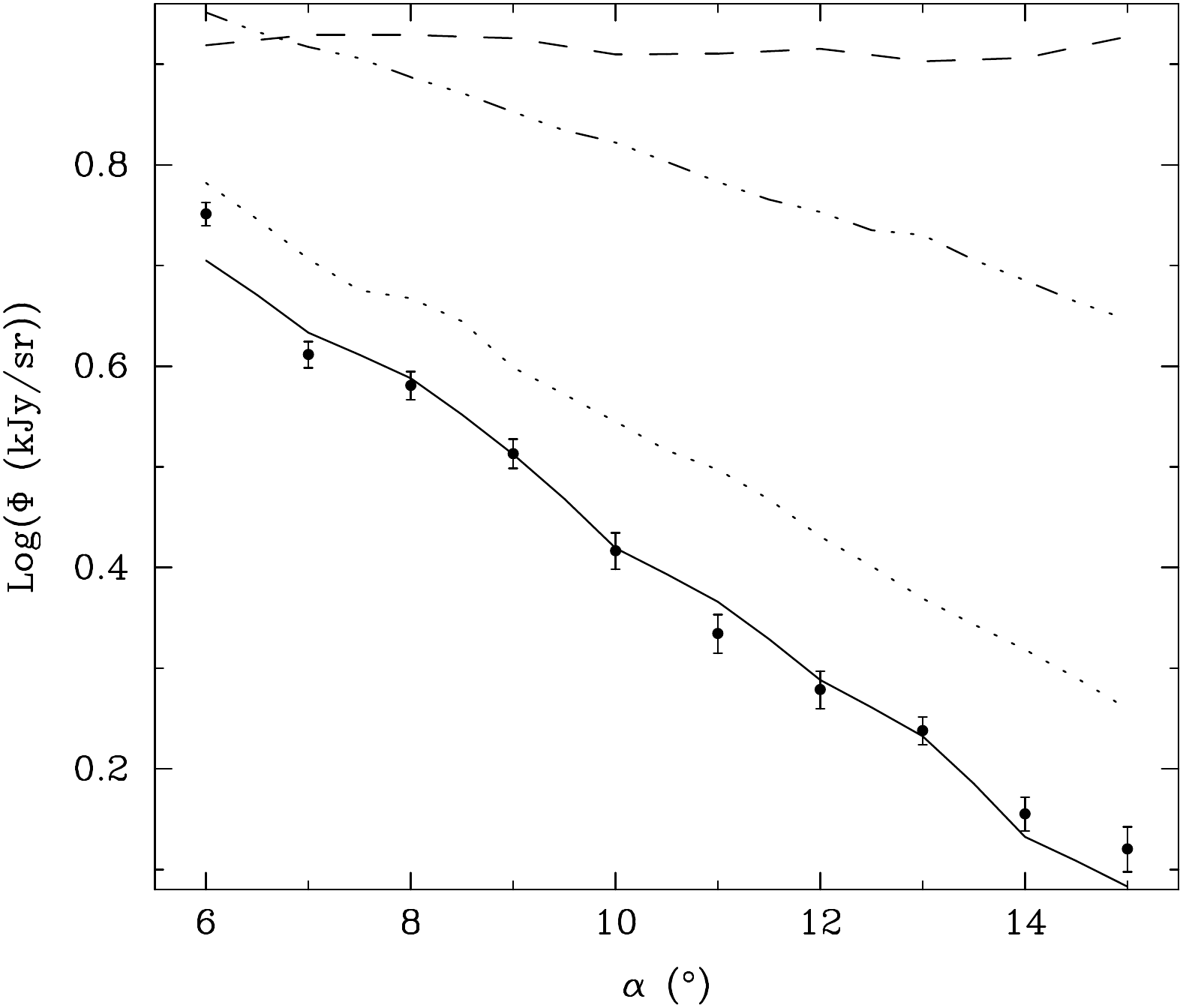}
\end{array}$
\end{center}
\caption{Angular flux profile $\Phi$ versus angle from the Galactic center $\alpha$. The diagram on the left is for the Gaussian pulsar distribution and the diagram on the right is for the pulsar distribution with no pulsars at the Galactic center. From top to bottom, the lines are the $\psi=0\degree$ (dashed), $\psi=30\degree$ (dash-triple dotted), $\psi=60\degree$ (dotted), and $\psi=0\degree$ (solid) plot. The plots are scaled with the $\psi=90\degree$ plot normalized to the flux of the WMAP haze. The points are the WMAP haze with errors from Ref.~\cite{Hooper07}.\label{pulsarprofile}}
\end{figure*}
In order to calculate the diffuse synchrotron flux $\Phi$ coming from pulsars, it is necessary to solve for the rate of $e^{+}e^{-}$ produced per units energy, volume, and time. This can be separated into two parts, the rate of electron (and positron) production in a given pulsar, $d^{2}N/\left(dE dt\right)$, and the spatial distribution of the pulsars in the Galaxy $\rho\left(\vec{x}\right)$.

For the number of electrons and positrons produced at a given energy in a given time, we use a model which does a fit to gamma-ray production from Galactic pulsars to determine the $e^{+}e^{-}$ rate from pulsars in the Galaxy~\cite{Zhang01}. Above about 1.5 GeV, this model is similar to the positron model in Ref.~\cite{Harding87}. For the combined $e^{+}e^{-}$ production rate, this model gives
\begin{eqnarray}
\frac{d^{2}N}{dE dt}&=&\left(5.7\times 10^{40}\rm\ GeV^{-1}s^{-1}\right) \\
&&\times f_{e}\dot{N}_{100}\left(\frac{E}{\rm GeV}\right)^{-1.6}\exp\left[-\frac{E}{80\rm\ GeV}\right]\ , \nonumber
\end{eqnarray}
where $f_{e}$ is the ratio of $e^{+}e^{-}$ to photon production in mature pulsars and $\dot{N}_{100}$ is the rate of pulsar formation in the Galaxy per hundred years. The free parameters $f_{e}$ and $\dot{N}_{100}$ are to account for the fact that the fit to gamma-ray pulsar data was done assuming a pulsar production rate of one per hundred years, so the overall normalization is unknown.

The distribution of pulsars in the Galaxy can be broken into two parts, a radial component $\rho_{s}\left(s\right)$ and a component against the Galactic plane $\rho_{z}\left(z\right)$. The full pulsar density is, then, $\rho\left(\vec{x}\right)=\rho_{s}\left(s\right)\rho_{z}\left(z\right)$. The perpendicular distribution $\rho_{z}$ should be peaked about the Galactic plane to match pulsar surveys with $\rho_{z}\left(-z\right)=\rho_{z}\left(z\right)$. This has been parameterized as~\cite{Narayan87}
\begin{equation}
\rho_{z}\left(z\right)=\frac{1}{0.61\pi^{1/2}}\exp\left[-\left(\frac{z}{0.61\rm\ kpc}\right)^{2}\right]\rm\ kpc^{-1}
\end{equation}
with the normalization $\int_{-\infty}^{\infty}dz=1$.

Because of observational bias in pulsar surveys, there is much model dependence in the shape of the radial distribution of pulsars $\rho_{s}$. For example, one must be careful in selection effects due to bright pulsars being more easily observable than dim ones. Also, in the inner part of the Galaxy, the pulsar distribution is largely unknown due to scattering and obscuration, so a targeted high-frequency radio survey would be needed to detect these innermost pulsars~\cite{Bailes92}.

This ambiguity in models has lead to two distinct fits to the pulsar survey data: a Gaussian fit with a peak at the Galactic center or a Boltzmann-like distribution which goes to zero at the Galactic center. As evidence for the latter, the Effelsberg 5 GHz Galactic center survey has not found any pulsars towards the Galactic center, but it is possible that any survey below 10 GHz cannot avoid scattering and obscuration effects~\cite{Lorimer04,Kramer00}. If a large number of pulsars in the inner kiloparsecs are observed by the Fermi Gamma-Ray Space Telescope, it could distinguish between these two models as well~\cite{Abdo09}.

The zero central density distribution is one in which the function peaks toward the Galactic center and then goes to zero at the center itself. This model is plausible if the pulsar distribution is similar to the distribution of supernova remnants and some astrophysical gas populations, which are in rings peaked about 4 kpc from the Galactic center~\cite{Bailes92, Stecker76, Leahy89, Robinson84}. This model of the pulsar radial distribution can be parameterized as~\cite{Lorimer04}
\begin{equation} \label{boltzmanndist}
\rho_{s}^{\rm 0GC}\left(s\right)=\left(376\right)^{-1}\left(\frac{s}{\rm kpc}\right)^{2.35}\exp\left[-\frac{s}{1.53\rm\ kpc}\right]\rm\ kpc^{-2}\ ,
\end{equation}
where $s$ is the distance from the Galactic center in the Galactic plane (in kpc). This distribution has been normalized to unity using $2\pi\int_{0}^{\infty}\rho_{s}(s)sds=1$.

Another radial distribution of Galactic pulsars is Gaussian, with a peak at the Galactic center falling off in the outer Galaxy. This model is motivated by stellar evolution arguments where the distribution is expected to be more densely peaked near the Galactic center, following the stellar distribution in the Galaxy. Such a radial distribution has been parameterized as~\cite{Narayan87}
\begin{equation} \label{gaussiandist}
\rho_{s}^{\rm Gauss}\left(s\right)=\frac{1}{64\pi}\exp\left[-\left(\frac{s}{8\rm\ kpc}\right)^{2}\right]\rm\ kpc^{-2}\ ,
\end{equation}
where $s$ is the distance from the Galactic center in the Galactic plane. This distribution has been normalized to unity as the zero central density distribution was. An exponential fit like the one used by Ref.~\cite{Kaplinghat09} can be modified to be indistinguishable from the Gaussian form~\cite{Narayan87}.
\subsection{Source Function for Dark Matter Annihilations}
To compare the pulsar source scenario with dark matter ones, we consider two different Navarro-Frenk-White (NFW)~\cite{Navarro97} type dark matter profiles with WIMP annihilations. In this model, the relativistic $e^{+}e^{-}$ are formed when dark matter particles annihilate in the Galactic dark matter halo. This can happen if the dark matter is its own antiparticle, as in the case of supersymmetric neutralinos~\cite{Finkbeiner07}. For the halo shape, we use a generalization of the standard NFW profile~\cite{Navarro97}. The generalization can be written in the form
\begin{equation}
\rho=\frac{\rho_{0}}{(r/r_{s})^{\gamma}(1+r/r_{s})^{3-\gamma}}\ ,
\end{equation}
for scale density $\rho_{0}$ and scale radius $r_{s}$. The original NFW profile has $\gamma=1$. A slightly steeper profile with $\gamma=1.2$ is also consistent with Milky-Way constraints~\cite{Klypin01,Battaglia05,Catena09}. We take $r_{s}=25\rm\ kpc$ as out canonical scale radius. The scale density is normalized to local earth dark matter density ($\sim 0.3\rm\ GeV/cm^{3}$)~\cite{PDG}. A standard WIMP has a weak-interaction cross-section of $\langle\sigma v\rangle=3\times 10^{-26}\rm\ cm^{3}s^{-1}$~\cite{Hooper07,Lavalle08}. For simplicity, we consider the direct annihilation channel where the product is exactly one positron and one electron with energy equal to the dark matter particle mass. We use 100 GeV dark matter particles as a canonical value. The morphological tests here are not sensitive to this choice. The source function for $e^{+}e^{-}$ formed from such a dark matter annihilation is
\begin{equation}
Q(E,\vec{x})^{\rm DM}=\frac{\rho^{2}}{2M^{2}}\langle\sigma v\rangle 2\delta(E-M)\ .
\end{equation}
\section{Synchrotron Flux from Sources}
\subsection{Flux from Pulsars}
\begin{figure*}[t]
\begin{center}$
\begin{array}{cc}
\includegraphics[width=3.2in]{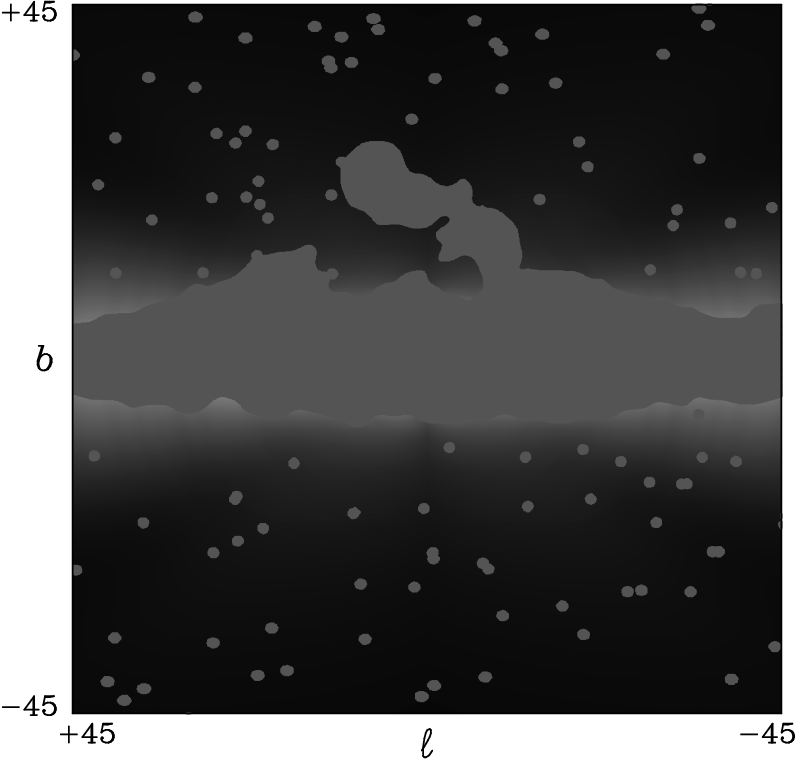} &
\hspace{1.0cm}
\includegraphics[width=3.2in]{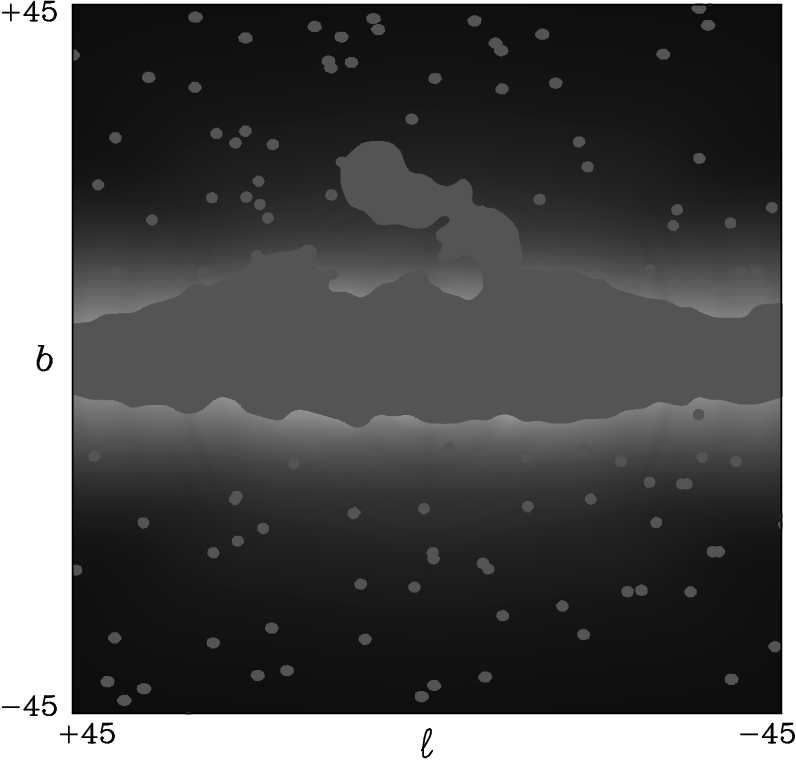}
\end{array}$
\end{center}
\caption{Relative flux $\Phi$ versus Galactic coordinates $\ell$ and $b$ for the Gaussian pulsar profile (left) and the pulsar profile with no pulsars at the Galactic center (right). The mask is that from WMAP K-band mask Kp4~\cite{Bennett03}.
\label{shadepulsar}}
\end{figure*}
For the standard pulsar distribution which increases toward the Galactic center, the $e^{+}e^{-}$ source term is 
\begin{equation}
Q(E,\vec{x})=\frac{d^{2}N}{dE dt}\rho_{z}(z)\rho_{s}^{\rm Gauss}(s)\ . 
\end{equation}
Using this with an overall scaling factor for the normalization of the particle production rate we find that the angular flux profile ($\Phi$ vs. $\alpha$) away from the Galactic plane fits the WMAP haze well, in agreement with Kaplinghat et al.~\cite{Kaplinghat09}. As in that work, we use an overall scaling factor to account for the uncertainties in particle production as well as a constant offset to account for uncertainties in background subtraction. We find that a physically reasonable scaling factor with electron production with similar magnitude to photon production in pulsars fits the data well.

We find that if one measures at an angle not orthogonal to the Galactic plane, but in an arbitrary direction, there is a strong dependence in the angular flux profile. As the direction of the line-of-sight shifts away from the vertical, there is a significant flattening in the angular flux profile (Fig.~\ref{pulsarprofile}). This leveling-off is due to the large scale with which the pulsar distribution drops off in the $s$-coordinate as compared to the relative steepness with which it drops off in the $z$-coordinate. Also, the overall magnitude of $\Phi$ changes by a factor of four to five at angle $\alpha=10\degree$. This effect is due to the fact that all fluxes should be equivalent in the Galactic center and begin dropping off with varying steepness at that angle. The 2-dimensional flux profile is shown in Fig.~\ref{shadepulsar}.

For the zero central density pulsar distribution, the source term becomes 
\begin{equation}
Q(E,\vec{x})=\frac{d^{2}N}{dE dt}\rho_{z}(z)\rho_{s}^{\rm 0GC}(s)\ .
\end{equation}
With this profile, we need a slightly different normalization to match the flux of the haze, which could be accounted for by modifying diffusion-loss parameters. In this case, however, the flattening-out effect happens at angles much closer to the normal to the Galactic plane. Because this profile has its maximum density at around 4 kpc out from the Galactic center, the line-of-sight integral keeps increasing out to $\alpha\approx 25\degree$. This creates a second peak in the angular flux profile at $\alpha\approx 20\degree$ before the flux begins to fall off at large angles. This can best be seen in the ($\ell,b$) diagram, Fig.~\ref{shadepulsar}, around $(\pm 20\degree,0\degree)$. There the magnitude increases slightly; this feature is not apparent in the WMAP haze. Unless this feature is detected, it is improbable that such a Galactic pulsar distribution could cause the haze. This could be seen in upcoming Planck observations which will likely have a smaller mask~\cite{Leach08}.
\subsection{Flux from Dark Matter Annihilations}
Here we consider Dark Matter annihilations as a source for the haze. Using the source term $Q(E,\vec{x})^{\rm DM}$ and normalizing the flux $\Phi$ away from the Galactic plane to the haze, we solve the diffusion-loss equation to determine the angular flux profile for the $\gamma=1.0$ and $\gamma=1.2$ NFW-type profiles. 

\begin{figure}[t]
\centering
\includegraphics[width=3.2in]{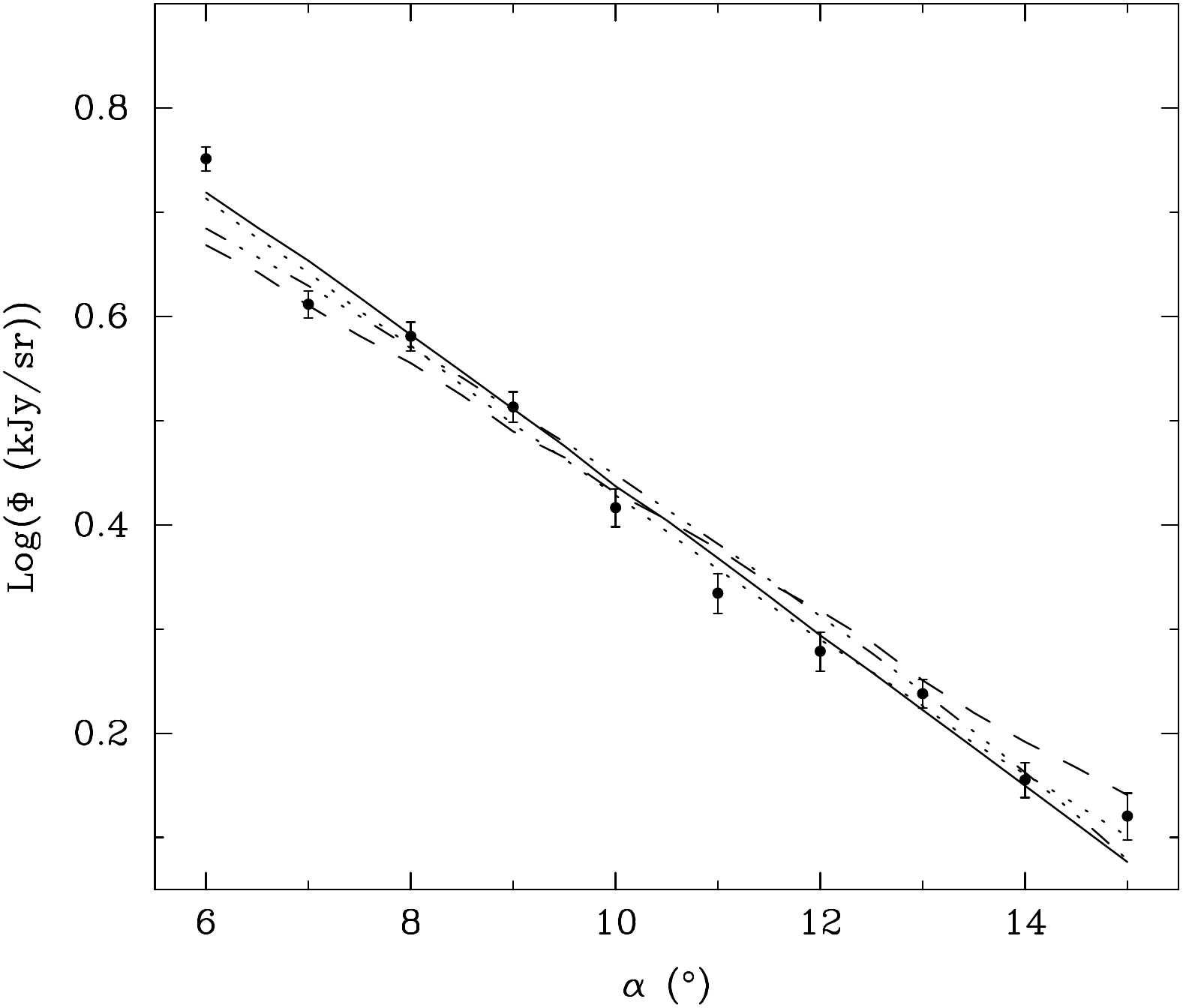}
\caption{Angular flux profile, $\Phi$ versus angle from the Galactic center $\alpha$ for the $\gamma=1.0$ and $\gamma=1.2$ dark matter profiles. The upper lines are for the $\psi=90\degree$ (solid) and $\psi=0\degree$ (dotted) curves using $\gamma=1.2$. The lower lines are for the $\psi=90\degree$ (dash-triple dotted) and $\psi=0\degree$ (dashed) curves using $\gamma=1.0$. The plots are scaled with the $\psi=90\degree$ plot normalized to the flux of the WMAP haze. The points are the WMAP haze with errors from Ref.~\cite{Hooper07}.\label{DMprofile}}
\end{figure}

The results of the angular flux profile for these cases are shown in Fig.~\ref{DMprofile}. As with the flux profiles due to pulsars, we used a constant offset to account for uncertainties in background subtraction as well as a boost factor to account for the possible clumpiness of the dark matter. We found that it only required reasonable boost factors of five to ten to match the haze. Due to the significant lack of dependence on direction, we have only plotted the $\psi=0\degree$ and $\psi=90\degree$ ($\Phi,\alpha$) curves for each case. Significantly, the turning-over effect seen with pulsars is not seen here. Towards the Galactic plane, there is a slight flattening of the curve, but it is not as drastic as with the pulsar source. Given proper foreground subtraction, this distinction may be used to distinguish between the dark matter and pulsar explanations of the WMAP haze.
\begin{figure*}[t]
\begin{center}$
\begin{array}{cc}
\includegraphics[width=3.2in]{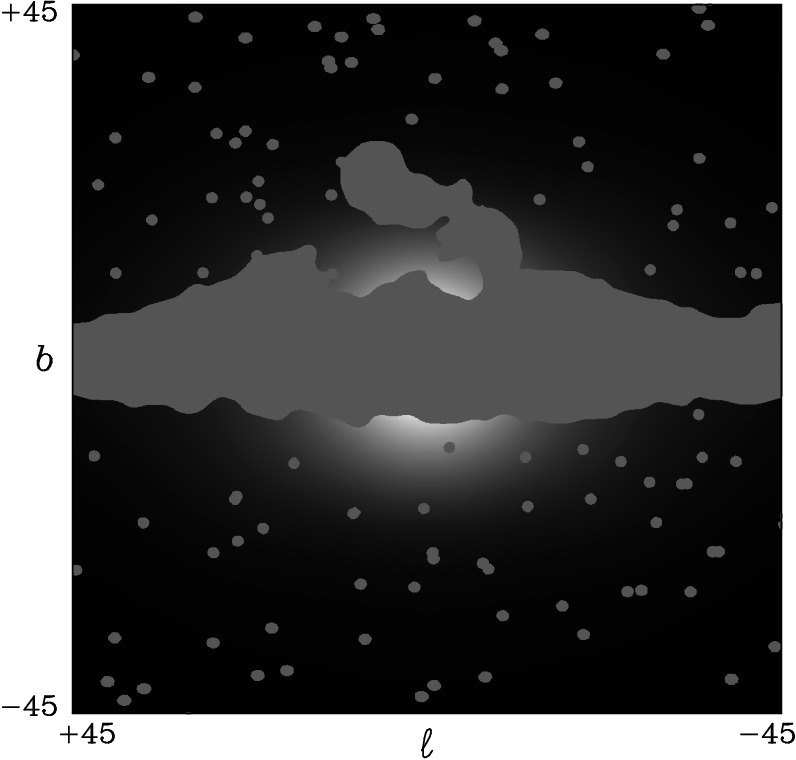} &
\hspace{1.0cm}
\includegraphics[width=3.2in]{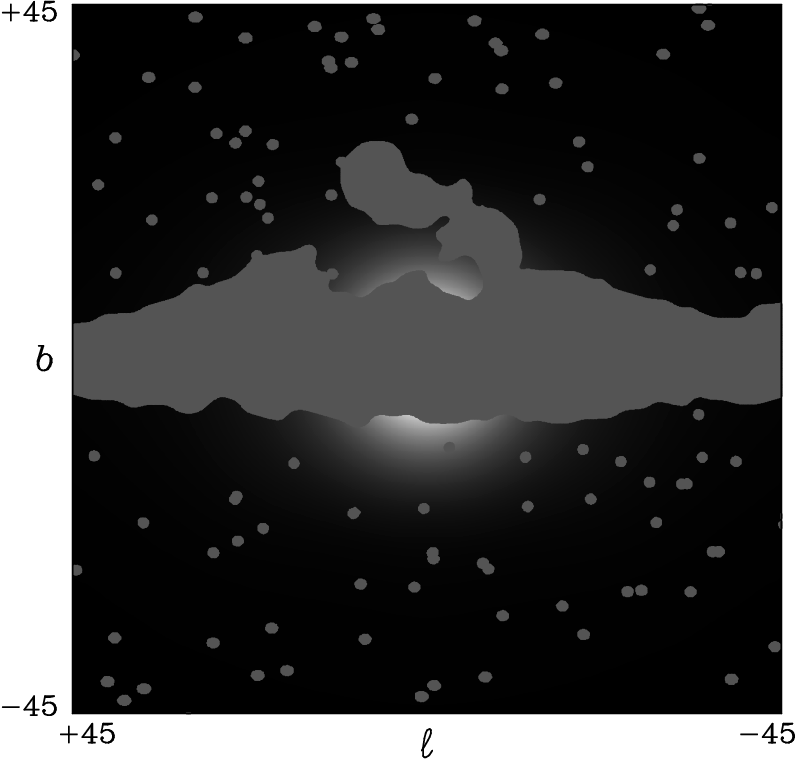}
\end{array}$
\end{center}
\caption{Relative flux $\Phi$ versus Galactic coordinates $\ell$ and $b$ for the standard NFW dark matter profile with $\gamma=1.0$ (left) and the NFW-type dark matter profile with $\gamma=1.2$ (right). The mask is that from WMAP K-band mask Kp4~\cite{Bennett03}.
\label{shadeDM}}
\end{figure*}
In the diagrams of relative magnitude of $\Phi$ versus $\ell$ and $b$, Fig.~\ref{shadeDM}, the two dark matter profiles have very similar characteristics. Both are significantly closer to spherical symmetry than the pulsar diagrams. There is some broadening in the $b$-direction, due to the size of the diffusion region being finite away from the Galactic plane. The cuspier $\gamma=1.2$ profile loses magnitude much more quickly away from the Galactic center, so it appears slightly more spherically symmetric on a plot weighted by magnitude. However, both fall off in similar elliptical regions. This sensitivity of the WMAP haze to the dark matter density profile can be used to constrain the Galactic halo's profile if the dark matter model is verified. In particular, this should be seen in Planck observations using a likely smaller Galactic mask~\cite{Leach08}.
\subsection{Dark Matter versus Pulsars}
Not only can the details of the pulsar or dark matter distributions be tested separately, but these tests can distinguish between the two models and determine the source of the WMAP haze. For comparison, we will use the best-fitting candidate from each category: the Gaussian pulsar distribution and the $\gamma=1.2$ dark matter distribution. Within the known values for each source, both can create a synchrotron signal large enough to cause the haze. Both sources have increased flux toward the Galactic center which falls off quickly away from it. 
\begin{figure}[ht]
\centering
\includegraphics[width=3.2in]{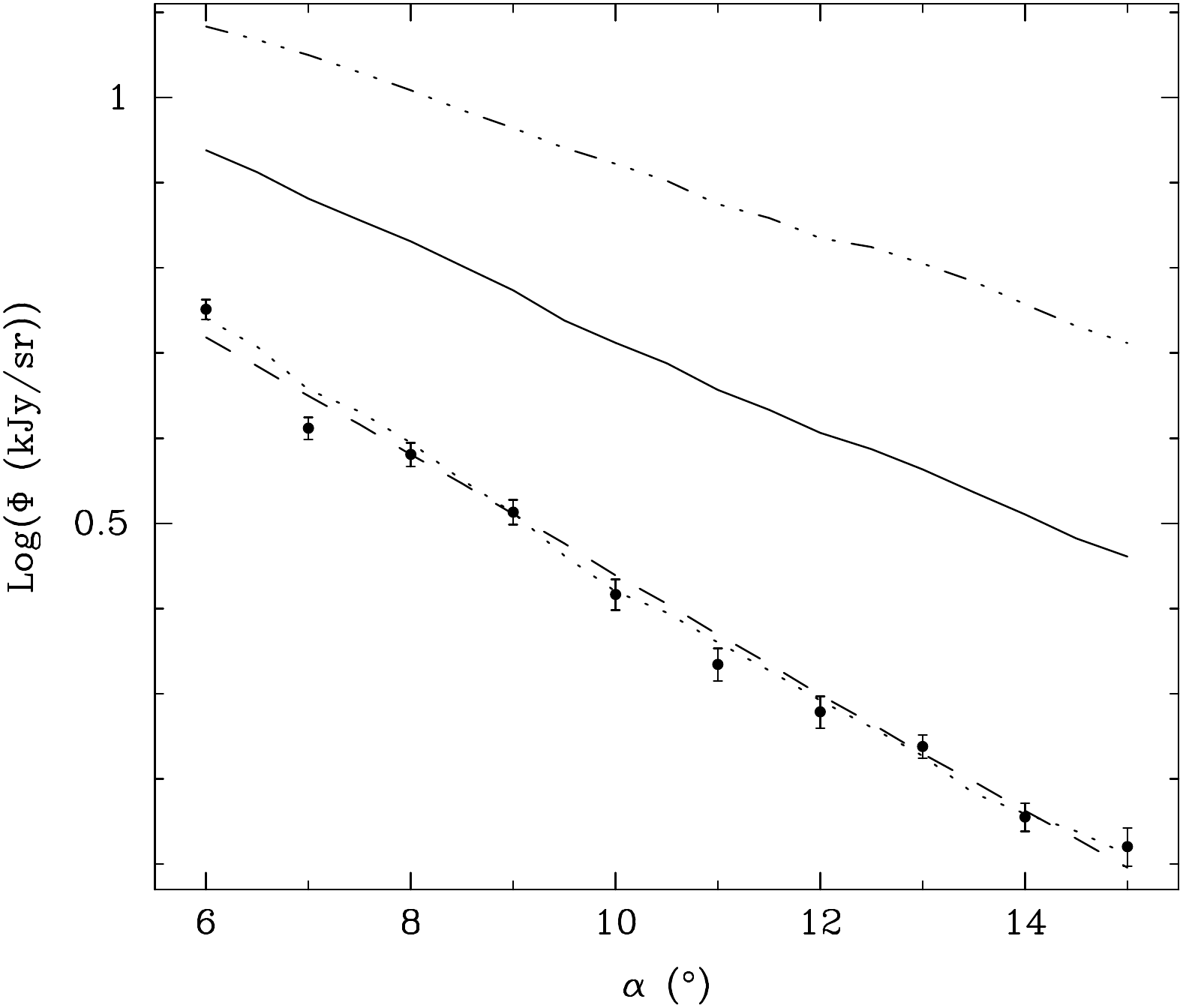}
\caption{Angular flux profile, $\Phi$ versus angle from the Galactic center $\alpha$ for the $\gamma=1.2$ dark matter profile and the Gaussian pulsar profile. The dashed line is the average over $\psi=30\degree-90\degree$ for the $\gamma=1.2$ dark matter distribution. The upper (dash-triple dotted) line and lower (dotted) line are the $\psi=30\degree$ and $\psi=90\degree$ curves for the Gaussian pulsar distribution, respectively. The solid curve is an average from $\psi=30\degree-90\degree$ for the Gaussian pulsar distribution. The plots are scaled with the $\psi=90\degree$ plot normalized to the flux of the haze. The points are the WMAP haze with errors from Ref.~\cite{Hooper07}.}
\label{fig:DMvspulsars}
\end{figure}

The largest difference between these two possible sources is the strength of the signal towards the Galactic plane. For the pulsar source, the strongest signal comes along the Galactic plane and does not decrease very quickly near it. The dark matter source has approximately equal signal strength in all directions, though it does lose strength more slowly along the Galactic plane than in other directions. Unfortunately, the Galactic plane itself masks most of this primary difference, though there are regions near the Galactic center which remain unmasked. We therefore compare the signal away from the Galactic plane.

As can be seen in Fig.~\ref{fig:DMvspulsars}, there are two significant differences between the pulsar source and the dark matter source when looking above the Galactic plane. For $\psi\ge 30\degree$, there is little distinguishable difference between the dark matter signal in any direction, so it is simply represented by the average over angles above 30 degrees. Primarily, the pulsar source has a different magnitude at different angles $\psi$. This can be seen by the vertical shift in the plots for different $\psi$ values. Even among the angles above the mask, there is a factor of 3 decrease in the flux $\Phi$ for lines-of-sight away from the Galactic plane. Secondarily, there is a flattening in the pulsar signal that is not present in the dark matter signal. Over the angles $\alpha=6\degree$ to $15\degree$, the $\psi=90\degree$ plot changes by roughly 50 percent more than the $\psi=30\degree$ plot. Such an effect is seen in the dark matter plots as well, but below the 10 percent level, which changes the signal minimally. Even above the mask, the two sources have substantial differences. With the Planck observatory's smaller expected Galactic mask, these differences should be even more apparent.
\section{Conclusions}
High-energy electrons/positrons moving in the Galactic magnetic field could be the source of the WMAP haze. Using the diffusion-loss equation, we have taken a set of models of likely sources and calculated the resulting synchrotron signal as a test for the models. The simplified equation can be solved analytically using a Green's function approach, where it is useful for an understanding of the underlying physics. We considered tests of several models of the haze source, namely a peaked Gaussian distribution of pulsars, a distribution of pulsars peaked away from the Galactic center, and two dark matter distributions as possible sources, all consistent with other constraints.

We find that the WMAP haze could be caused by diffuse synchrotron emission due to pulsars, in agreement with Kaplinghat et al.~\cite{Kaplinghat09}. Moreover, we find that if the haze is caused by pulsars, the angular flux profile from the Galactic center should be peaked more sharply when using a line-of-sight away from the Galactic plane and is flattened significantly when using a line-of-sight closer to the Galactic plane. Also, if the pulsar distribution vanishes in the Galactic center, a brightness about $20\degree$ out along the Galactic plane should be visible. 

With annihilating dark matter, the angular flux profile from Galactic center is largely independent of direction and has rough spherical symmetry. If the signal is found to be strongly spherically symmetric, then this would be an indication that annihilating dark matter could be the true source. 

This directional-dependence of the angular flux profile would be a smoking gun for pulsars causing the haze as opposed to the more spherical dark matter explanation. The Planck probe's increased sensitivity, larger number of bands, enhanced models, and expected smaller Galactic mask will test these models~\cite{Leach08,Planck06,Stolyarov01}. This will be instrumental in determining the source of the WMAP haze as an astrophysical signal or the indirect detection of the dark matter. 
\begin{acknowledgments}
We thank Z. Chacko, Doug Finkbeiner, Ted Jacobson, and Manoj Kaplinghat for useful discussions. JPH is partially supported by the UMD/GSFC Joint Institute for Space Studies and the Maryland Center for Fundamental Physics. KNA is partially supported by NSF Theoretical Physics grant No. 0757966.
\end{acknowledgments}
\bibliography{bibliography}
\end{document}